\documentclass{article}
\usepackage{authblk}
\usepackage{amsmath,amsfonts,amssymb}
\usepackage{txfonts}
\usepackage[T1]{fontenc}
\usepackage{hyperref}

\def\PL#1{Phys.\ Lett.\ {\bf#1}}
\def\PRL#1{Phys.\ Rev.\ Lett.\ {\bf#1}}
\def\PR#1{Phys.\ Rev.\ {\bf#1}}
\def\NP#1{Nucl.\ Phys.\ {\bf#1}}

\def\JoP#1{J.\ Phys.\ {\bf#1}} \def\IJMP#1{Int.\ J. Mod.\ Phys.\ {\bf #1}}
\def\MPL#1{Mod.\ Phys.\ Lett.\ {\bf #1}}

\def\JHEP#1{JHEP\ {\bf#1}}

\def\EPJ#1{Eur.\ Phys.\ J.\ {\bf#1}}
\def\nc{noncommutative }\def\adj{{\rm ad}}
\def\bd{\begin{displaymath}}\def\ed{\end{displaymath}}
\def\be{\begin{equation}}\def\ee{\end{equation}}
\def\bea{\begin{eqnarray}}\def\eea{\end{eqnarray}}
\def\ba{\begin{array}}\def\ea{\end{array}}

\newcommand{\arxiv}[1]{\href{https://arxiv.org/abs/#1}{arXiv:#1}}
\newcommand{\bibx}[3]{#1, ``#2'', \arxiv{#3}}
\newcommand{\bibxp}[4]{#1, ``#2'', #3, \arxiv{#4}}
\newcommand{\bibp}[3]{#1, ``#2'', #3}

\begin{document}
\begin{titlepage}
\title{$\kappa$-deformed phase spaces, Jordanian twists, Lorentz-Weyl algebra and dispersion relations}
\vskip80pt
\author[1]{D. Meljanac{\footnote{Daniel.Meljanac@irb.hr}}}
\author[2]{S. Meljanac{\footnote{meljanac@irb.hr}}}
\author[3,4]{S. Mignemi{\footnote{smignemi@unica.it}}}
\author[5]{R. \v Strajn{\footnote{rina.strajn@unidu.hr}}}
\affil[1]{Division of Materials Physics, Ru\dj er Bo\v skovi\'c Institute, Bijeni\v cka cesta 54, 10002 Zagreb, Croatia}
\affil[2]{Division of Theoretical Physics, Ru\dj er Bo\v skovi\'c Institute, Bijeni\v cka cesta 54, 10002 Zagreb, Croatia}
\affil[3]{Dipartimento di Matematica e Informatica, Universit\`{a} di Cagliari, viale Merello 92, 09123 Cagliari, Italy}
\affil[4]{INFN, Sezione di Cagliari, Cittadella Universitaria, 09042 Monserrato, Italy}
\affil[5]{Department of Electrical Engineering and Computing, University of Dubrovnik, \'{C}ira Cari\'{c}a 4, 20000 Dubrovnik, Croatia}
\date{}
\renewcommand\Affilfont{\itshape\small}
\clearpage\maketitle
\thispagestyle{empty}

\begin{abstract}
We consider $\kappa$-deformed relativistic quantum phase space and possible implementations of the Lorentz algebra. There are two ways of performing such implementations. One is a simple extension where the Poincar\'e algebra is unaltered, while the other is a general extension where the Poincar\'e algebra is deformed. As an example we fix the Jordanian twist and the corresponding realization of noncommutative coordinates, coproduct of momenta and addition of momenta. An extension with a one-parameter family of realizations of the Lorentz generators, dilatation and momenta closing the Poincar\'e-Weyl algebra is considered. The corresponding physical interpretation depends on the way the Lorentz algebra is implemented in phase space. We show how the spectrum of the relativistic hydrogen atom depends on the realization of the generators of the Poincar\'e-Weyl algebra.
\end{abstract}

\end{titlepage}

\section{Introduction}

A major open problem in modern theoretical physics is the incompatibility of quantum mechanics and general relativity. In the search of a successful theory of quantum gravity, the non-renormalizability of perturbative general relativity suggests that it might be necessary to abandon the assumption of a continuous spacetime and introduce a minimal length $l$, which is expected to be of the order of the Planck length, namely $1,6\times 10^{-35}$ m, with an associated momentum cutoff of the order of the Planck mass. 

In order for the minimal length and the momentum cutoff to be compatible with Lorentz invariance, the corresponding deformation scale should be included into the commutation relations, implying a deformation of the Lorentz symmetry and of the dispersion relations of particles. These considerations lead to the introduction of noncommutative spacetimes and of their associated symmetry algebras, which are deformations of the classical symmetries of general relativity \cite{Snyder,Moyal,Connes,DFR}. A suitable mathematical framework for investigating their properties is that of Hopf algebras \cite {Majid,ref8,AADD}. 

Among noncommutative spacetimes, one of  the most extensively studied is the $\kappa$-Minkowski spacetime, with its symmetry algebra, the $\kappa$-Poincar\'{e} Hopf algebra
\cite{LNR,Ruegg}. The $\kappa$-Minkowski spacetime is a Lie algebra-type deformation of ordinary Minkowski space and is symmetric under a $\kappa$-deformation of the Poincar\'{e} algebra, which can be endowed with a Hopf algebra structure. The $\kappa$-Poincar\'e algebra has a well defined meaning with 10 generators, without dilatation, defined modulo choice of basis. It is known that $\kappa$-deformed phase space is obtained with a so-called Heisenberg double construction from dual pair of quantum $\kappa$-Poincar\'e algebra and quantum $\kappa$-Poincar\'e group \cite{nu1,nu2,nu3,nu4,LSW}. This construction is kind of reference point for construction of $\kappa$-deformed phase spaces.

In spite of the fact that the Planck scale is not directly accessible, it might nevertheless be possible to indirectly study the physical effects of the previous assumptions. Phenomenological investigations of this kind are usually carried out in the framework of Doubly Special Relativity (DSR) \cite{Amelino1,Mag}, which is strictly related to the $\kappa$-Poincar\'e formalism, in particular to the deformation of the Lorentz symmetry. DSR theories mainly investigate the consequences of the deformation of the dispersion relations for particles. In particular, in the case of photons, the deformation may imply an energy-dependent speed of light.  Some proposals have been advanced based on the cumulative effect of the energy-dependent velocity of photons traveling in quantum spacetime across cosmological distances, that could cause a measurable delay in the arrival time of  gamma-ray bursts from distant galaxies \cite{AmelinoGRB}. Other proposals for the detection of quantum-gravity effects rely on ground-based experiments like Holometer in Fermilab \cite{Holo} and various quantum optics experiments \cite{QO1, QO2}.

Different potentially observable effects have also been discussed, like modifications of the uncertainty relations, corrections to the energy spectrum of the harmonic oscillator or of the hydrogen atom, etc.~\cite{phenomenology}, but their detection is presently out of reach for experiments, and hence they have only a theoretical interest.

Noncommutative spaces can be described as deformed phase spaces generated by \nc coordinates $\hat x_\mu$ and commutative momenta $p_\mu$ obeying $[p_\mu,p_\nu]=0$,
as, for example, the above mentioned $\kappa$-Minkowski space, which is discussed in Sec. 2. For given commutation relations $[\hat x^\mu,\hat x^\nu]$, there is freedom in the choice of the remaining 
commutation relations, $[p_\mu,\hat x^\nu]=-i\varphi_\mu{}^\nu (p)$, with $\varphi_\mu{}^\nu$ functions of $p$, such that $\varphi_{\mu\nu}$ goes to the Minkowskian metric $\eta_{\mu\nu}$ when the deformation parameter, the Planck mass, goes to infinity. Once these are fixed, there is a unique realization of \nc coordinates of type $\hat{x}^\mu =x^\alpha \varphi_\alpha{}^\mu(p)$, where $x^\mu$ and $p_\mu$ generate the ordinary Heisenberg algebra.
Moreover, for a given realization of $\hat x_\mu$ there exists a 
unique star product $f(x)\star g(x)$ and a corresponding twist operator $\mathcal{F}$, up to a freedom in the ideal related to the star product. The coproduct $\Delta p_\mu$ and the addition of momenta are also fixed.
In general, the realization of \nc coordinates  $\hat x_\mu$, the twist $\mathcal{F}$, the coproduct $\Delta p_\mu$ and the star product  $f(x)\star g(x)$ are uniquely interrelated \cite{MMMP,MMPP}.

In order to extend the deformed phase space generated by coordinates $\hat x_\mu$ and $p_\mu$ to include the action of a  Lorentz algebra there are two possibilities. One is the simple extension in which the 
momenta $p_\mu$ transform as vectors under Lorentz transformations generated by $M_{\mu\nu}$. The other is  a more general extension in which the momenta transform in a deformed
way \cite{MMMP}. In fact, we can define new canonical coordinates $P_\mu$ and $X_\mu$ and Lorentz generators $M_{\mu\nu}$ such that
\bea
&&P_\mu=Sp_\mu S^{-1},\qquad X_\mu=Sx_\mu S^{-1},\cr
&&M_{\mu\nu}=S(x_\mu p_\nu-x_\nu p_\mu)S^{-1}=X_\mu P_\nu-X_\nu P_\mu,
\eea
where 
\be
S=e^{ix_\alpha\Sigma_\alpha(p)}.
\ee
Then,
\bea
&&P_\mu=e^{i\,\adj_{x_\alpha}\Sigma_\alpha}(p_\mu),\cr
&&X_\mu=e^{i\,\adj_{x_\alpha}\Sigma_\alpha}(x_\mu).
\eea
We note that $M_{\mu\nu}$ and $p_\mu$ generate a deformed Poincar\'e algebra, while $M_{\mu\nu}$ and $P_\mu$ generate the standard  Poincar\'e algebra, since $P_\mu$ 
transforms as a vector under the action of $M_{\mu\nu}$. The wave equations are essentially determined by writing $P^2(p)$ in terms of $p^2$ and $v\cdot p$, where $v_\mu$ is defined in Sec. 2, see Secs. 3, 4, 5. 

It is important to remark that the implementation
of the Poincar\'e algebra generated by $M_{\mu\nu}$ and $P_\mu$  into the deformed phase space is arbitrary. There is no physical principle that fixes $P_\mu(p)$,
 i.e.~$\Sigma(p)$. The theory of Hopf algebras, the Drinfeld twist, and quantum Lie algebras are inspirational for physics, but without implementing physical principles they cannot predict
physical results. However, there are claims that a given twist uniquely determines observables and dispersion relations \cite{Asc1703}, but the corresponding construction seems rather ad hoc.

One simple family of possible implementations of the Poincar\'e-Weyl algebra is given in Sec. 4.1. Another interesting implementation related to the natural realization of the 
$\kappa$-Poincar\'e algebra is discussed in  Sec. 4.2.

In the present paper we fix a specific realization $\hat x^\mu$, with associated Jordanian twist,  star product and coproduct, and show that the physical results depend on the implementation of the Lorentz algebra, labelled by a parameter $u$, leading to a family of Poincar\'e-Weyl algebras. A different example of implementation of the Poincar\'e algebra is the
$\kappa$-Poincar\'e algebra \cite{LNR}. Therefore, we show that there exists freedom in the implementation of the Lorentz algebra in $\kappa$-Minkowski space.

The plan of the paper is as follows: in Sec. 2 we describe $\kappa$-deformed relativistic quantum phase space. A simple extension with Lorentz algebra is given in Sec. 3, together 
with some explicit examples. In Sec. 4 we present a general extension with Lorentz algebra leading to a deformed Poincar\'e algebra. It includes examples of interpolation between left- and right-covariant realizations corresponding to the Poincar\'e-Weyl algebra, as well as the natural realization of $\kappa$-Poincar\'e algebra. Dispersion relations and physical consequences are discussed in Sec. 5, where it is shown how the spectrum of the relativistic hydrogen atom depends on the representation of the Lorentz algebra. Conclusions are presented in Sec. 6.


\section{$\kappa$-deformed relativistic quantum phase space}

The $\kappa$-Minkowski spacetime \cite{LNR} is generated by noncommutative coordinates $\hat{x}_\mu$, with $\mu =0,1,...,n-1$, which satisfy Lie-algebra type commutation relations
\begin{equation}
[\hat{x}_i,\hat{x}_j]=0, \quad [\hat{x}_0,\hat{x}_i]= ia_0 \hat{x}_i,
\end{equation}
where $a_0 \propto 1/\kappa$ is a deformation parameter assumed to be the order of the Planck length. These commutation relations can be written in a covariant way
\begin{equation} \label{kappaM}
[\hat{x}^\mu, \hat{x}^\nu] = iC^{\mu\nu}{}_\lambda \hat x^\lambda = i(a^\mu \hat{x}^\nu -a^\nu \hat{x}^\mu ),
\end{equation}
where $C^{\mu\nu}{}_\lambda = a^\mu \delta^\nu_\lambda - a^\nu \delta^\mu_\lambda$ is a structure constant, $a^\mu =\frac{1}{\kappa} v^\mu$ and $v^\mu$ is a vector with $v^2=v^\mu v^\nu \eta_{\mu\nu} \in \{ -1,0,1\}$, where $\eta_{\mu\nu}=diag(-1,1,...,1)$ is the metric of the Minkowski space.

The momenta $p_\mu$ are commutative and satisfy
\begin{equation}
[p_\mu,p_\nu]=0, \quad [p_\mu, \hat{x}^\nu]=-i\varphi_\mu{}^\nu (p).
\end{equation}
The noncommutative coordinates $\hat{x}^\mu$ can then be realized as\footnote{One can consider a more general realization of the form $\hat{x}^\mu =x^\alpha \varphi_\alpha{}^\mu(p) + \chi^\mu(p)$, but in this paper, only the cases with $\chi^\mu(p)=0$ will be considered.} \cite{MMMP,MSt,K-JMSt,MK-J,smdmSS,JPA1110}
\begin{equation}\label{xreal}
\hat{x}^\mu =x^\alpha \varphi_\alpha{}^\mu(p),
\end{equation} 
where $x^\mu$ is the canonical coordinate conjugate to the momentum $p_\mu$ and $x^\mu$ and $p_\mu$ satisfy the undeformed Heisenberg algebra
\begin{equation}
[x^\mu,x^\nu]=0, \quad [p_\mu,p_\nu]=0, \quad [p_\mu,x^\nu]=-i\delta_\mu^\nu.
\end{equation}
From \eqref{kappaM}, it follows that the functions $\varphi_{\mu\nu}(p)$ satisfy
\begin{equation} \label{eq za fi}
\frac{\partial \varphi_\sigma{}^\mu}{\partial p_\alpha} \varphi_\alpha{}^\nu -\frac{\partial \varphi_\sigma{}^\nu}{\partial p_\alpha} \varphi_\alpha{}^\mu = C^{\mu\nu}{}_\alpha \varphi_\sigma{}^\alpha,
\end{equation}
where $C^{\mu\nu}{}_\alpha =a^\mu \delta^\nu_\alpha - a^\nu \delta^\mu_\alpha$. The differential equation \eqref{eq za fi} has infinitely many solutions, which are connected by similarity transformations. For a given solution $\varphi_\mu{}^\nu (p)$ there exists a unique star product $f(x)\star g(x)$ which is noncommutative and associative. The star product can be written using the corresponding twist operator $\mathcal{F}$
\begin{equation}
f(x)\star g(x) =m\left[\mathcal{F}^{-1}(\triangleright \otimes \triangleright) (f(x) \otimes g(x)) \right], \quad f(x), g(x) \in \mathcal{A},
\end{equation}
where $m$ is the multiplication map and $\mathcal{A}$ is the algebra generated by $x_\mu$. Moreover, $\triangleright:\mathcal H \otimes \mathcal A \to \mathcal A$ is the action defined by $f(x)\triangleright g(x) = f(x)g(x)$ and $p_\mu\triangleright g(x)=-i\frac{\partial g(x)}{\partial x^\mu}$.  A twist $\mathcal{F}$ is an invertible bidifferential operator that satisfies the cocycle and normalization conditions
\begin{align}
& (\mathcal{F} \otimes 1) (\Delta_0 \otimes id) \mathcal{F}= (1\otimes \mathcal{F})(id \otimes \Delta_0) \mathcal{F}, \\
& m(id \otimes \epsilon )\mathcal{F} = m(\epsilon \otimes id) \mathcal{F} =1,
\end{align}
where $\epsilon$ is the counit and $\Delta_0$ is the undeformed coproduct. The twist can be constructed from $\varphi_{\mu\nu}(p)$ in the Hopf algebroid approach \cite{GGHMM, JMS, JMPS,JMP,MMP}, and can be transformed to a twist in the Hopf algebra approach.


Let us consider a few examples. Realizations related to a simple interpolation between Jordanian twists \cite{MMPP} are
\begin{equation} \label{x-hat-u}
\hat x_\mu
 = (x^{(u)}_\mu - (1-u)a_\mu (x^{(u)}\cdot p^{(u)}))(1-uA^{(u)}), 
\end{equation}
where $A^{(u)}=-a^\alpha p^{(u)}_\alpha \equiv -a\cdot p^{(u)}$. The corresponding Drinfeld twists are given by
\begin{equation}
\mathcal F_{(u)}^{-1}= e^{-\Delta_0(uD^{(u)}A^{(u)})}
e^{\ln(1+A^{(u)})\otimes D^{(u)}}
e^{u(D^{(u)}A^{(u)}\otimes 1 + 1 \otimes A^{(u)}D^{(u)})},
\end{equation}
where $D^{(u)}=x^{(u)}\cdot p^{(u)}$ is the dilatation operator.
The superscripts $(u)$ appearing in the previous formulae are explained in Sec. 4. They essentially refer to the fact that different values of $u$ correspond to different parametrizations of the canonical phase space.
The commutation relations involving the noncommutative coordinates \eqref{x-hat-u} are
\begin{eqnarray}\label{comrel}
&&[\hat x^\mu,\hat x^\nu]=i(a^\mu\hat x^\nu-a^\nu\hat x^\mu),\cr
&&[p_\mu,\hat x^\nu]=-i\left(\delta^\mu_\nu-a^\nu(1-u)p_\mu\right)(1-uA).
\end{eqnarray}

The left covariant realization \cite{K-JMSt}, for $u=1$, is given by
\begin{equation}\label{real-left}
\hat x_\mu = x^L_\mu(1-A^L)=x_\mu(1+a\cdot p^L)=x_\mu Z^{-1},
\end{equation}
where $A^L=-a\cdot p^L$ and $Z^{-1}=1+a\cdot p^L$. The corresponding Jordanian twist is given by
\begin{equation}\label{twist-left}
\mathcal F_L^{-1} = e^{D^L\otimes \ln(1-A^L)}.
\end{equation}

The right covariant realization \cite{K-JMSt}, for $u=0$, is given by
\begin{equation}\label{real-right}
\hat x_\mu = x^R_\mu -a_\mu(x^R \cdot p^R),
\end{equation}
and in the case of timelike $a_\mu$ reduces to the well-known Magueijo-Smolin model \cite{Mag}.
The corresponding Jordanian twist is given by
\begin{equation}\label{twist-right}
\mathcal F_R^{-1} = e^{\ln(1+A^R)\otimes D^R}.
\end{equation}


For the natural realization \cite{K-JMSt,MK-J} (classical basis), the realization of noncommutative coordinates is given by
\begin{equation}\label{natural}
\hat x_\mu = x^N_\mu Z^{-1}-(a\cdot x^N)p^N_\mu,
\end{equation}
where $Z$ is the shift operator, defined by
\begin{equation}
[Z, \hat x^\mu] = ia^\mu Z,
\end{equation}
that takes the explicit form
\begin{equation}\label{zeta}
Z^{-1} = \sqrt{1+a^2(p^{N})^2} +a\cdot p^{N}.
\end{equation}

The corresponding twist does not satisfy the cocycle condition in the Hopf algebra sense, but it does in the Hopf algebroid sense
\cite{JMS, JMSalgebroid, JKMalgebroid,MSS}.
Generally, for noncommutative coordinates \eqref{xreal}, the corresponding twist in the Hopf algebroid approach is given by \cite{MMPP, MS}
\begin{equation}
\mathcal F^{-1} = e^{-ip_\alpha \otimes x^\alpha}e^{ip^W_\beta \otimes \hat x^\beta},
\end{equation}
where $p^W_\mu$ is related to the Weyl realization and is defined by
\begin{equation}
[p^W_\mu, \hat x^\nu] = -i\left( \frac{\mathcal C}{1-e^{-\mathcal C}} \right){}^\nu{}_\mu.
\end{equation}
where $\mathcal C^\mu{}_\nu=C^{\mu\alpha}{}_\nu p^W_\alpha$. There is a one-to-one correspondence between $e^{ik\cdot \hat x}$ and $e^{ik\cdot x^W}$. The relation between $p_\mu$ and $p^W_\mu$ is given in \cite{MMMP,MMPP,JMPS}.

For $\kappa$-Minkowski space, $\mathcal C^\mu{}_\nu = a^\mu p_\nu - (a\cdot p)\delta^\mu_\nu$, and the commutator $[p^W_\mu, \hat x^\nu]$ is given by \cite{K-JMSt,JPA1110,JMPS}
\begin{equation} \label{pW}
[p^W_\mu, \hat x^\nu] =
-i\delta_\mu^\nu \frac{A^W}{e^{A^W}-1}
-i a^\nu p^W_\mu \frac{e^{A^W}-1-A^W}{(e^{A^W}-1)A^W}.
\end{equation}


\section{Simple extension with Lorentz algebra}

The Lorentz algebra is given by the commutation relations
\begin{equation} \label{Lorentz}
[M_{\mu\nu}, M_{\rho\sigma}] = i(\eta_{\mu\rho} M_{\nu\sigma}- \eta_{\mu\sigma} M_{\nu\rho} +\eta_{\nu\rho} M_{\mu\sigma} -\eta_{\nu\sigma} M_{\mu\rho}).
\end{equation}
If one includes also the generators of translation $p_\mu$ and the dilatation operator $D$, the resulting algebra is the Poincar\'{e}-Weyl algebra with the remaining commutation relations given by
\begin{align}
[D, M_{\mu\nu}]&=0, \label{uc-DM}\\
[p_\mu,p_\nu]&=0, \label{uc-pp}\\
[M_{\mu\nu},p_\lambda]&=i( \eta_{\mu\lambda} p_\nu -\eta_{\nu\lambda} p_\mu),  \label{uc-Mp}\\
[D,p_\mu]&=ip_\mu.\label{uc-Dp}
\end{align}
Requiring that the commutation relations of the Poincar\'{e}-Weyl algebra \eqref{Lorentz}-\eqref{uc-Dp} remain undeformed, $M_{\mu\nu}$ and $D$ can be realized in terms of the ordinary phase space variables in the usual way
\begin{equation}
M_{\mu\nu}=x_\mu p_\nu-x_\nu p_\mu, \quad D=x\cdot p \equiv x^\alpha p_\alpha.
\end{equation}
The commutation relations between the  Lorentz generators $M_{\mu\nu}$ and dilatation $D$ with noncommutative coordinates $\hat x^\mu$ are:
\begin{align}
\label{c-Mhx}
\begin{split}
[M_{\mu\nu}, \hat x_\lambda] &=-i(x_\mu \varphi_{\nu\lambda} - x_\nu \varphi_{\mu\lambda})
+ix^\alpha\left(
\frac{\partial \varphi_{\alpha\lambda}}{\partial p^\mu} p_\nu
-
\frac{\partial \varphi_{\alpha\lambda}}{\partial p^\nu} p_\mu
\right), \\
&=
-i\hat x^\gamma (\varphi^{-1}_{\gamma\mu} \varphi_{\nu\lambda} - \varphi^{-1}_{\gamma\nu} \varphi_{\mu\lambda})
+i\hat x_\gamma (\varphi^{-1})^{\gamma\alpha}\left(
\frac{\partial \varphi_{\alpha\lambda}}{\partial p^\mu} p_\nu
-
\frac{\partial \varphi_{\alpha\lambda}}{\partial p^\nu} p_\mu
\right),
\end{split}\\
\label{c-Dhx}
\begin{split}
[D, \hat x_\lambda] &= -i\hat x_\lambda + i x^\alpha p_\beta \frac{\partial \varphi_{\alpha\lambda}}{\partial p_\beta} \\
&=-i\hat x_\lambda + i \hat x_\gamma (\varphi^{-1})^{\gamma\alpha} p_\beta \frac{\partial \varphi_{\alpha\lambda}}{\partial p_\beta}.
\end{split}
\end{align}
The commutation relations \eqref{c-Mhx} and \eqref{c-Dhx} depend on the realization $\hat x^\mu = x^\alpha \varphi_\alpha{}^\mu(p)$ and all Jacobi identities are satisfied.

The coalgebra sector can be obtained using the twist operator
\begin{equation} \label{Delta g preko F}
\Delta g =\mathcal{F} \Delta_0 g \mathcal{F}^{-1}, \quad g \in \{ p_\mu, M_{\mu\nu}, D\},
\end{equation}
where $\Delta_0 g =g\otimes 1+1 \otimes g$ and the antipodes are
\begin{equation}
S(g) = \chi S_0(g) \chi^{-1},
\end{equation}
where $\chi = m[(S_0\otimes1)\mathcal F]$.


For the left covariant realization, we find \cite{K-JMSt,MK-J,JPA1110}
\begin{equation}\begin{split}
[M^L_{\mu\nu}, \hat x_\lambda]&=
-i(x^L_\mu \eta_{\nu\lambda} - x^L_\nu \eta_{\mu\lambda})(1-A^L)
-ix^L_\lambda(a_\mu p^L_\nu - a_\nu p^L_\mu)
\\&=
-i(\hat x_\mu \eta_{\nu\lambda} - \hat x_\nu \eta_{\mu\lambda})
-i \hat x_\lambda \frac{a_\mu p^L_\nu - a_\nu p^L_\mu}{1-A^L},
\end{split}\end{equation}
and from the corresponding Jordanian twist \cite{BorPac, BuKim}, the coalgebra structure of the $\kappa$-deformed Poincar\'e-Weyl algebra is
\begin{align}
\Delta p^L_\mu &= p^L_\mu \otimes (1-A^L) + 1\otimes p^L_\mu,\label{deltap-L} \\
\Delta M^L_{\mu\nu} &= M^L_{\mu\nu} \otimes 1 + 1\otimes M^L_{\mu\nu} - D^L \otimes \frac{a_\mu p^L_\nu - a_\nu p^L_\mu}{1-A^L}, \\
\Delta D^L &= D^L \otimes \frac1{1-A^L} + 1\otimes D^L,
\end{align}
while the antipodes are
\begin{align}
S(p^L_\mu) &= \frac{-p^L_\mu}{1-A^L}, \\
S(M^L_{\mu\nu}) &= -M^L_{\mu\nu}-D^L (a_\mu p^L_\nu - a_\nu p^L_\mu) 
, \\
S(D^L) &= -D^L (1-A^L).
\end{align}


For the right covariant realization, we find \cite{K-JMSt,MK-J,JPA1110}
\begin{equation}\begin{split}
[M^R_{\mu\nu}, \hat x_\lambda]&=
-i(x^R_\mu \eta_{\nu\lambda} - x^R_\nu \eta_{\mu\lambda})
\\&=
-i\left[\hat x_\mu \eta_{\nu\lambda} - \hat x_\nu \eta_{\mu\lambda}
+(a_\mu\eta_{\nu\lambda} - a_\nu\eta_{\mu\lambda})(\hat x \cdot p^R) \frac1{1+A^R}\right],
\end{split}\end{equation}
and from the corresponding Jordanian twist, the coalgebra structure of the $\kappa$-deformed Poincar\'e-Weyl algebra is
\begin{align}
\Delta p^R_\mu &= p^R_\mu \otimes 1 + (1+A^R)\otimes p^R_\mu, \\
\Delta M^R_{\mu\nu} &= M^R_{\mu\nu} \otimes 1 + 1\otimes M^R_{\mu\nu}+
\frac{a_\mu p_\nu^R-a_\nu p_\mu^R}{1+A^R} \otimes D^R \\
\Delta D^R &= D^R \otimes 1 + \frac1{1+A^R}\otimes D^R.
\end{align}
The antipodes are
\begin{align}
S(p^R_\mu) &= \frac{-p^R_\mu}{1+A^R}, \\
S(M^R_{\mu\nu}) &= -M^R_{\mu\nu}+(a_\mu p^R_\nu - a_\nu p^R_\mu)
D^R
, \\
S(D^R) &= -(1+A^R)D^R .
\end{align}


For the natural realization \cite{K-JMSt,MK-J,JPA1110,MSS,plb12}, we get the $\kappa$-Poincar\'e algebra \cite{LNR}. The coproducts are given by
\begin{align}
\Delta p_\mu^N = p_\mu^N\otimes Z^{-1} +1\otimes p_\mu^N -a_\mu p_\alpha^L Z\otimes (p^N)^\alpha, \\
\Delta M_{\mu\nu}^N = M_{\mu\nu}^N \otimes 1+1\otimes M_{\mu\nu}^N -a_\mu (p^L)^\alpha Z\otimes M_{\alpha\nu}^N +a_\nu (p^L)^\alpha Z
\otimes M_{\alpha\mu}^N,
\end{align}
and the antipodes are
\begin{align}
S(p_\mu^N) &=(-p_\mu^N -a_\mu p^L\cdot p^N)Z, \\
S(M_{\mu\nu}^N) &= -M_{\mu\nu}^N -a_\mu  (p^L)^\alpha M_{\alpha\nu}^N +a_\nu  (p^L)^\alpha M_{\alpha\mu}^N.
\end{align}


\section{General extensions with Lorentz algebra}

In general, for a given twist, it is possible to leave the Lorentz sector \eqref{Lorentz} undeformed but allow a deformation of the commutators \eqref{uc-Mp} and \eqref{uc-Dp}
\begin{align}
[M_{\mu\nu}, p_\lambda] &= \Gamma_{\mu\nu\lambda}(p) ,\\
[D, p_\lambda] &= \sigma_\lambda(p),
\end{align}
such that all Jacobi identities are satisfied. Hence, the physical consequences for observables and dispersion relations are not uniquely determined by a given twist or star product, but also  depend on the implementation of the Lorentz algebra. In other words, a given twist does not fix the implementation of the Lorentz algebra \cite{MMMP,MSS,plb12,Loret}. Here we demonstrate this observation.

\subsection{Interpolation between the left and right covariant realization}
Let us start with the left covariant realization \eqref{real-left} and the corresponding twist \eqref{twist-left} and consider the following one-parameter similarity transformations related to $x^L$, $p^L$. In the following, we shall omit the label $L$ for simplicity.

The coordinates $x_\mu^{(u)}$ characterized by the parameter $u$ can be written in terms of coordinates $x_\mu$ ($u=1$) as
\begin{equation}\label{x-u}
x_\mu^{(u)} = \text{e}^{-(1-u)iDA}\, x_\mu\, \text{e}^{(1-u)iDA} = (x_\mu + (1-u)a_\mu (x \cdot p))(1 - (1-u)A).
\end{equation}
The corresponding momenta are
\begin{equation}\label{p-u}
p_\mu^{(u)} = \text{e}^{-(1-u)iDA}\, p_\mu\, \text{e}^{(1-u)iDA} =\frac{p_\mu}{1-(1-u)A}.
\end{equation}
The left covariant realization is reproduced for $u=1$ and the right one for $u=0$.


The generators $M^{(u)}_{\mu\nu}$ and $D^{(u)}$ can also be expressed in terms of the ones corresponding to the left covariant realization
\begin{align} \label{M-u}
M^{(u)}_{\mu\nu}& = \text{e}^{-(1-u)iDA}\, M_{\mu\nu}\, \text{e}^{(1-u)iDA} = M_{\mu\nu} + (1-u)D(a_\mu p_\nu - a_\nu p_\mu),\\
D^{(u)}&= \text{e}^{-(1-u)iDA}\, D_\mu\, \text{e}^{(1-u)iDA} = D(1-(1-u)A).\label{nova}
\end{align}
The transformed momenta $p^{(u)}_\mu$, Lorentz generators $M^{(u)}_{\mu\nu}$ and dilatation $D^{(u)}$ generate the Poincar\'e-Weyl algebra 
eqs.~\eqref{Lorentz}-\eqref{uc-Dp}. However, the algebra generated by $p_\mu$, $M^{(u)}_{\mu\nu}$ and $D^{(u)}$ is a deformed Poincar\'e-Weyl algebra
\begin{align}\label{poincare}
[M^{(u)}_{\mu\nu}, p_\lambda] &=i( \eta_{\mu\lambda} p_\nu -\eta_{\nu\lambda} p_\mu)
+ i(1-u)p_\lambda(a_\mu p_\nu - a_\nu p_\mu) , \\
[D^{(u)}, p_\mu] &= ip_\mu(1-(1-u)A).\label{Du-pL}
\end{align}

The inverse relations to \eqref{x-u}-\eqref{nova} are given by
\begin{align}
\label{xxu}
x_\mu &= (x^{(u)}_\mu - (1-u)a_\mu (x^{(u)} \cdot p^{(u)}))(1+(1-u)A^{(u)}), \\
\label{ppu}
p_\mu &= \frac{p_\mu^{(u)}}{1 + (1 - u)A^{(u)}}, \\
\label{MMu}
M_{\mu\nu} &= M^{(u)}_{\mu\nu} - (1-u)D^{(u)}(a_\mu p^{(u)}_\nu - a_\nu p^{(u)}_\mu), \\
\label{DDu}
D&=D^{(u)}(1+(1-u)A^{(u)}).
\end{align}
The shift operator can also be written in terms of the interpolating variables
\begin{equation}
Z=\frac{1+(1-u)A^{(u)}}{1-uA^{(u)}}.
\end{equation}
The noncommutative coordinates written in terms of the interpolating variables are given by \eqref{x-hat-u}.
Using relations \eqref{p-u}-\eqref{DDu}, the homomorphism property of the coproduct and the known coproducts \cite{MMPP}, one can then find the coproducts of $p^{(u)}_\mu$, $M^{(u)}_{\mu\nu}$ and $D^{(u)}$,
\begin{align}
\label{Delta-pu}
\Delta p^{(u)}_\mu &=
\frac
{p^{(u)}_\mu\otimes(1-uA^{(u)}) + (1+(1-u)A^{(u)})\otimes p^{(u)}_\mu}
{1\otimes1 + u(1-u)A^{(u)}\otimes A^{(u)}}, \\
\begin{split}\label{Delta-Mu}
\Delta M^{(u)}_{\mu\nu} &= M^{(u)}_{\mu\nu} \otimes 1 + 1 \otimes M^{(u)}_{\mu\nu}
-u D^{(u)}(1+(1-u)A^{(u)})\otimes \frac{a_\mu p^{(u)}_\nu - a_\nu p^{(u)}_\mu}{1- uA^{(u)}}
\\
&+(1-u)\frac{a_\mu p^{(u)}_\nu -a_\nu p^{(u)}_\mu }{1+(1-u)A^{(u)}} \otimes D^{(u)} (1-uA^{(u)})
\end{split}\\ \label{Delta-Du}
\Delta D^{(u)} &= \left(
\frac1{1+(1-u)A^{(u)}}\otimes D^{(u)} + D^{(u)} \otimes \frac1{1-uA^{(u)}}
\right)
(1\otimes 1 + u(1-u) A^{(u)}\otimes A^{(u)}).
\end{align}


The antipodes of $p^{(u)}_\mu$, $M^{(u)}_{\mu\nu}$ and $D^{(u)}$ are given by
\begin{align}
\label{S-pu}
S(p_\mu^{(u)})&=\frac{-p^{(u)}_\mu}{1+(1-2u)A^{(u)}}, \\
\label{S-Mu}
S(M_{\mu\nu}^{(u)})&=-M_{\mu\nu} - uD(a_\mu p^{(u)}_\nu - a_\mu p^{(u)}_\nu)
+(1-u)\frac{a_\mu p^{(u)}_\nu - a_\mu p^{(u)}_\nu}{1-uA^{(u)}}D^{(u)}(1-uA^{(u)}), \\
\label{S-Du}
S(D^{(u)})&=-\frac{1+(1-2u)A^{(u)}}{1-uA^{(u)}} D^{(u)} (1-uA^{(u)}).
\end{align}
Equations \eqref{Delta-pu}-\eqref{S-Du} for coproducts $\Delta p_\mu^{(u)}$, $\Delta M^{(u)}_{\mu\nu}$, $\Delta D^{(u)}$ and antipodes $S(p^{(u)})$, $S(M^{(u)}_{\mu\nu})$, $S(D^{(u)})$ define a one-parameter family of $\kappa$-deformed Poincar\'e-Weyl Hopf algebras.

Starting from the right covariant realization, instead of the left one, one arrives at the same results. One can also define a Hopf algebroid structure \cite{LSW,MS,MSSt}.

\subsection{Relation between natural and left covariant realizations}

The so-called natural realization \cite{MSt,K-JMSt,MK-J,JPA1110} of the $\kappa$-Minkowski spacetime is defined by \eqref{natural}
It does not belong to the family of realizations parametrized by $u$, but it can nevertheless be related to the left covariant realization
\begin{equation}\label{nat1}
x^{N}_\mu= x_\mu +a\cdot x\frac{p_\mu -\frac{a_\mu }{2} p^2Z}{1+\frac{a^2}{2} p^2Z},
\end{equation}
where 
\begin{equation}\label{nat2}
Z^{-1}=1+a\cdot p=\sqrt{1+a^2(p^N)^2}+a\cdot p^N.
\end{equation}
The relations between the generators of the Poincar\'{e} algebra are given by
\begin{align}
p^{N}_\mu &= p_\mu -\frac{a_\mu}{2} p^2 Z, \label{pN preko L} \\
M^{N}_{\mu\nu} &= M_{\mu\nu} -\frac{1}{2} (L_{\mu\alpha} a_\nu -L_{\nu\alpha} a_\mu) p^\alpha Z. \label{MN preko L}
\end{align}
where $L_{\mu\nu} =x_\mu p_\nu$. 

The coproducts of $p^{N}_\mu$ and $M^{N}_{\mu\nu}$ are known from the literature \cite{LNR,Ruegg}
\begin{align}
\Delta p^{N}_\mu &= p^{N}_\mu \otimes Z^{-1} + 1\otimes p^{N}_\mu -a_\mu p_\alpha^{N} Z\otimes (p^N)^\alpha + \frac{a_\mu }{2} \Box Z \otimes a\cdot p^{N}, \label{CoprodpN} \\
\Delta M^{N}_{\mu \nu} &=M_{\mu\nu}^{N} \otimes 1+1 \otimes M_{\mu\nu}^{N} -(a_\mu(p^L)^\alpha Z \otimes M_{\alpha\nu}^{N} -a_\nu (p^L)^\alpha
Z\otimes M_{\alpha\mu}^{N}), \label{CoprodMN}
\end{align}
where $\Box = -p^2Z =\frac{2}{a^2} (1-\sqrt{1+a^2(p^N)^2})$.

As in the previous subsection, it is easily shown that using the relations between the two realizations and the homomorphism property of the coproduct, one obtains the same expressions. However, a crucial point in showing this for the coproduct of $M_{\mu\nu}^{N}$ are the so-called tensor identities \cite{JMS,JMSalgebroid,JKMalgebroid,MSS}. They can be calculated from the undeformed ones, $\mathcal{R}_0 =x_\mu \otimes 1-1 \otimes x_\mu=0$, using the twist operator. For the left covariant realization, \eqref{twist-left} gives
\begin{equation} \label{id L}
\mathcal{F} \mathcal{R}_0 \mathcal{F}^{-1} =0 = x_\mu \otimes Z -1\otimes x_\mu +D\otimes a_\mu Z.
\end{equation}
The way to show that one obtains the same result is to first calculate the coproduct of $M^{N}_{\mu\nu}$ starting from \eqref{MN preko L}, using the homomorphism property \cite{JMS,JMSalgebroid,JKMalgebroid,MSS}. On the other hand, the known coproduct \eqref{CoprodMN} is written in terms of the left covariant realization variables using \eqref{pN preko L} and \eqref{MN preko L}. Finally, using the tensor identities \eqref{id L}, one finds that the expressions agree.

\section{Dispersion relations and physical consequences}

\def\la{\lambda}\def\a{\alpha}\def\d{\delta}
\def\f{\phi}\def\h{\theta}\def\i{\iota}
\def\k{\kappa}\def\q{\psi}\def\o{\omega}
\def\m{\mu}\def\n{\nu}\def\V{\varphi}
\def\D{\Delta}\def\L{\Lambda}
\def\de{\partial}\def\mo{{-1}}\def\ha{{1\over 2}}
\def\nn{\nonumber}\def\lb{\label}\def\bb{\bibitem}

\def\po{{p_0\over\k}}\def\Eo{{E\over\k}}\def\ar{{\a\over r}}\def\fact{\left(1-u{p_0\over\k}\right)}
\def\hx{{\hat x}}

In this section we consider some physical consequences of the deformations of the phase space discussed in the previous sections.
It must be noted that the physics depends on both the realization of the noncommutative coordinates $\hat{x}_\mu$ (in our case labelled with the parameter $u$) and on the specific representation chosen for the Poincar\'e algebra.
As discussed in Sec. 4, one may adopt a representation in which the algebra is unaltered, or one in which it is deformed. In the
following, we consider the latter possibility.\footnote{In the literature on DSR this is the usual assumption; notice however that in \cite{Asc1703} a different 
interpretation of the formalism has been given.} Here we start with a Jordanian twist $\mathcal{F}$, eq.\eqref{twist-left}, (the same twist as in ref. \cite{Asc1703}), the corresponding realization of noncommutative coordinates, $\hat{x}_\mu= x_\mu (1+a\cdot p)$ \eqref{real-left}, coproduct $\Delta p_\mu$ \eqref{deltap-L} and addition of momenta $(k\oplus q)_\mu =k_\mu(1+a\cdot q)+ q_\mu$ and we consider a 
one-parameter family of realizations of the Lorentz generators $M_{\mu\nu}^{(u)}$ \eqref{M-u} and dilatation $D^{(u)}$ \eqref{nova}, $M_{\mu\nu}^{(u)}= x_\mu p_\nu -x_\nu p_\mu +(1-u) x\cdot p (a_\mu p_\nu -a_\nu p_\mu)$, $D^{(u)} =x\cdot p(1+(1-u)a\cdot p)$. They satisfy commutation relations \eqref{poincare}-\eqref{Du-pL} with the momentum $p_\lambda$. The corresponding Casimir operator is $C=P^2\equiv(p^{(u)})^2=\frac{p^2}{(1+(1-u)a\cdot p)^2}$.

\subsection{Classical mechanics}
We start by studying the classical motion determined by the Hamilton equations with the deformed Poisson brackets corresponding to the classical limit
of the commutation relations \eqref{comrel}.
We shall limit our considerations to the case where $v^\m=(1,0,0,0)$, because this choice does not break the rotational invariance and is the most interesting
phenomenologically. For $u=0$ it corresponds to the well-known model introduced by Magueijo and Smolin in \cite{Mag}, in the realization originally proposed
by Granik \cite{Gr} and then investigated in several papers \cite{Mig,KMM,Gho}.

With the timelike choice of $v^\m$ considered in this section, the nonvanishing Poisson brackets derived from the commutators \eqref{comrel} are
\begin{align}\lb{pb}
\{\hx^0,\hx^i\}=-{\hx^i\over\k},\qquad\{p_0,\hx^0\}=-\left(1-(1-u){p_0\over\k}\right)\fact,\nn\\
\{p_i,\hx^0\}=(1-u){p_i\over\k}\fact,\qquad\{p_i,\hx^j\}=-\d_i^j\fact.
\end{align}

We choose a deformed relation between Lorentz generators $M_{\mu\nu}^{(u)}$ and momenta $p_\lambda$, as in \eqref{poincare}. Its Casimir invariant,
\be
C={-p_0^2+p_i^2\over\left[1-(1-u)\,a\cdot p\right]^2},
\ee
can be chosen as the Hamiltonian for a free particle. In our case, we have
\be\lb{ham}
H=\ha\,{-p_0^2+p_i^2\over\left[1-(1-u)\po\right]^2}=-{m^2\over2},
\ee
where $m$ is the Casimir mass.  From this expression follows that the rest energy of the classical particle is not $m$, but rather 
\be
m_0={m\over1+(1-u){m\over\k}}.
\ee

From the Poisson brackets \eqref{pb} and the Hamiltonian \eqref{ham}, one obtains the Hamilton equations
\be
\dot\hx_\m={p_\m\left(1+u\po\right)\over\left[1-(1-u)\po\right]^2},\qquad\dot p_\m=0.
\ee
It follows that for this class of models the classical 3-velocity of a free particle of arbitrary mass, defined as the phase velocity
${\bf v}^K_i={\dot\hx_i\over\dot\hx_0}$,
is equal to the relativistic one,
${\bf v}_i={p_i\over p_0}$, as was already noticed for the Magueijo-Smolin case in \cite{Gr,Mig}. In the interacting case, instead, the 
motion can differ from that of special relativity.

\subsection{Quantum mechanics}
We pass now to investigate the quantum theory. In this case, the velocity is usually identified with the group velocity
for wave propagation with dispersion relation \eqref{ham},
\be
{\bf v}^H_i={\de p_0\over\de p_i}={p_i\over p_0-(1-u){m^2\over\k}\left[1-(1-u)\po\right]},
\ee
For massive particles the group velocity differs from the phase velocity. The contrast between the classical and
quantum definition of velocity is a well-known problem in DSR theories \cite{vel}.
It has been shown however that, at least under some assumptions, the physical predictions coincide for both choices, if one takes into account the
deformation of translations and the effects of the relative locality of observers \cite{ALR}.

In any case, since massless particles have both phase and  group velocity equal to $c$, no effects of time delay in the detection of cosmological
photons, like those discussed in \cite{AmelinoGRB}, are predicted in the present models.

However, some nontrivial consequences follow from the lack of invariance of the Hamiltonian for $p_0\to-p_0$. 
For example the Klein-Gordon (KG) equation for a free particle can be written in the form
\be
\left[-p_0^2+p_i^2+m^2\left(1-(1-u)\po\right)^2\right]\f=0.
\ee

Adopting the Hilbert space realization of the operators following from \eqref{x-hat-u} and the discussion in Sec. 4,
\bea\lb{quant}
&&p_\m\to -i{\de\over\de x_\m},\qquad\hx_i\to x_i\left(1-{iu\over\k}{\de\over\de x_0}\right),\cr
&&\hx_0\to\left( x_0+{i(1-u)\over\k}\ x_i\,{\de\over\de x_\i}\,\right)\left(1-{iu\over\k}{\de\over\de x_0}\right),
\eea
the Klein-Gordon equation takes the form
\be
\left[{\de^2\over\de x_0^2}-{\de^2\over\de x_i^2}+m^2\left(1+{i(1-u)\over\k}{\de\over\de x_0}\right)^2\right]\f=0.
\ee

Its plane wave solutions are given by
\be\lb{planewave}
\q=e^{-i(\o^\pm x^0-k_ix^i)},
\ee
where 
\be
\o^\pm=-{(1-u)m^2/\k\pm\sqrt{k^2+m^2-(1-u)^2k^2m^2/\k^2}\over1-(1-u)^2m^2/\k^2}.
\ee
The two values of $\o^\pm$ correspond to positive and negative energy states and can be interpreted as belonging to particles and antiparticles
\cite{CM}.
They have different absolute value because the invariance of the Klein-Gordon equation for $p_0\to-p_0$ is broken.
In particular, it follows that the rest mass $m^-_0$ of antiparticles  differs from that of particles, $m^+_0$, namely,
\be
m^\pm_0={m\over1\mp(1-u){m\over\k}}.
\ee
Of course, a more rigorous treatment of this nontrivial property should be undertaken in the context of QFT.

Further physical effects may arise in more complex systems, where some interaction is present. Since the theories studied in this
paper are intrinsically relativistic, such effects must be sought in non-Newtonian systems: a relevant example is the relativistic hydrogen atom.
Therefore, we analyze the first-quantized  Klein-Gordon equation for a particle of Casimir mass $m$ and unit charge in a central electric field.
This can be considered as a first approximation to the relativistic hydrogen atom, when the fermionic nature of the electron is neglected.

We write the KG equation in the form
\be\lb{KG}
\left[-\left(p_0-\ar\right)^2+p_i^2+m^2\left(1-(1-u)\po\right)^2\right]\f=0,
\ee
where $\a$ is proportional to the central electric charge and $r=\sqrt{\hx_i^2}$, and make the substitutions \eqref{quant}.
We pass then to spherical coordinates, imposing the ansatz
\be\lb{ansatz}
\f(x_0,x_i)=\sum_{ml}e^{-iEx_0}Y_{lm}(\h,\V)\psi_{lm}(r),
\ee
where $Y_{lm}(\h,\V)$ are spherical harmonics. Expanding to first order in $1\over\k$, we obtain from \eqref{KG},
\bea
&&\bigg[{\de^2\over\de r^2}+\left(1-u\Eo\right){2\over r}{\de\over\de r}-\left(1-2u\Eo\right){l(l+1)-\a^2\over r^2}+\left(1-u\Eo\right){2\a E\over r}\cr
&&\quad+~E^2-m^2\left(1-2(1-u)\Eo\right)\bigg]\q_{lm}=0.
\eea

The regular solutions of this equation can be written as
\be\lb{solution}
\q_{lm}=e^{-\D r}r^{-\ha+u\Eo+\L}\,L^{2\L}_n(2\D r),
\ee
where 
\be\lb{D-L}
\D=\sqrt{m^2-E^2-2(1-u)\,m^2\Eo},\qquad\L=\sqrt{[(l+\ha)^2-\a^2](1-2u\Eo)-{u\over2}\Eo},
\ee 
 and $L^\a_n$ are generalized Laguerre  polynomials. Moreover, $n$ is an integer given by
\be\lb{eigen}
n=-\ha+{E\a(1-u\Eo)\over\D}-\L.
\ee

To zeroth order in $1/\k$, we obtain the energy eigenvalues of the relativistic hydrogen atom as \cite{Kleinert}
\be\lb{spec}
E_{(0)}^{nl}=\pm{(n-\ha-\la)\over\sqrt{(n-\ha-\la)^2+\a^2}}\ m=\pm {mN\over\sqrt{N^2+\a^2}},
\ee
where $\la=\sqrt{(l+\ha)^2-\a^2}$,\ $N=n-\ha-\la$, and the minus sign refers to the negative energy states (antiparticles).
The energy includes the rest mass. 

Expanding \eqref{eigen} around \eqref{spec}, one can obtain the corrections to the spectrum due to the quantum geometry,
\be
E^{nl}\sim E_{(0)}^{nl}\left[1+{1\over\k}\left({E_{(0)}\left(1-2u+u{E_{(0)}^2\over m^2}\right)}\mp{u(m^2-E_{(0)}^2)^{3/2}\over\a m^2}\
{l(l+1)-\a^2\over\sqrt{(l+\ha)^2-\a^2}}\right)\right]\nn
\ee
\be\lb{spect}
=E_{(0)}^{nl}\left[1\pm{mN\over\k\sqrt{N^2+a^2}}\left(1-u\,{N^2-2\a^2\over N^2+\a^2}\right)\mp{u\,\a^2\over(N^2+\a^2)}\
{l(l+1)-\a^2\over\sqrt{(l+\ha)^2-\a^2}}\right].
\ee
In particular, for the Magueijo-Smolin model \cite{Mag} , $u=0$,
\be
E^{nl}=\pm {mN\over\sqrt{N^2+\a^2}}\left(1\pm{1\over\k}{mN\over\sqrt{N^2+\a^2}}\right).
\ee
From \eqref{spect} it follows that the leading corrections to the relativistic spectrum are of order $m/\k\sim10^{-23}$.
As for the free KG equation, one  can observe a breaking of the particle/antiparticle symmetry, but even the spectrum of the positive energy
states is deformed.

\subsection{Left covariant realization and $\kappa$-Poincar\'e algebra}
We consider now the same problem for the left covariant realization corresponding to the $\kappa$-Poincar\'e  algebra \cite{LNR, Ruegg}, see Sec.~4.2,
using a representation of the Lorentz algebra generated by $M_{\mu\nu}$ in \eqref{MN preko L}.
As before we make the choice $a^\m=({1\over\k},0,0,0)$. 
From \eqref{nat1}, \eqref{nat2} and \eqref{pN preko L} or \eqref{real-left} and \eqref{natural} it follows that
\be\label{repnat}
\hx_\mu=x_\mu\left(1+{p_0\over\k}\right)=x^N_\m\, Z^{-1}-{1\over\k}x^N_0p^N_\m,
\ee
and the nonvanishing Poisson brackets in the classical limit become
\be
\{\hx^0,\hx^i\}=-{\hx^i\over\k},\qquad\{p_\m,\hx^\n\}=-\left(1+{p_0\over\k}\right)\d^\n_\m.
\ee
The free particle Hamiltonian can be chosen proportional to the Casimir operator corresponding to the Lorentz realization \eqref{MN preko L},
which is \cite{K-JMSt}
\be
H=\ha\ p^2Z=\ha\ {-p_0^2+p_i^2\over1+{p_0\over\k}}=-{m^2\over2},
\ee
with $m$ the Casimir mass. It follows that the rest energy is $m_0=m\left(-{1\over2\k}+\sqrt{1-{m^2\over4\k^2}}\right) $.

The Hamilton equations then read
\be
\dot\hx_0=-{p_0+{p_0^2\over2\k}+{p_i^2\over2\k}\over1+{p_0\over\k}},\qquad\dot\hx_i=p_i,\qquad\dot p_\m=0.
\ee
The classical 3-velocity is therefore given by 
\be
{\bf v}^K_i={\dot\hx_i\over\dot\hx_0}=-{p_i\left(1+{p_0\over\k}\right)\over p_0+{p_0^2\over2\k}+{p_i^2\over2\k}},
\ee
and coincides with ${\bf v}^H_i={\de p_0\over\de p_i}$.

We consider now quantum mechanics. The Klein-Gordon equation for a free particle can be written in the form
\be
\left[-p_0^2+p_i^2+m^2\left(1+\po\right)\right]\f=0.
\ee
From \eqref{repnat} follows that one can adopt the Hilbert space realization of the operators 
\be\lb{quantnat}
p_\m\to -i{\de\over\de x_\m},\qquad \hx_\m\to x_\m\left(1-{i\over\k}{\de\over\de x_0}\right).
\ee
The KG equation then becomes
\be
\left[-{\de^2\over\de x_0^2}+{\de^2\over\de x_i^2}+m^2\left(1-{i\over\k}{\de\over\de x_0}\right)\right]\f=0.
\ee
Its plane wave solution has the general form \eqref{planewave}, where now
\be
\o^\pm=-{m^2\over2\k}\pm\sqrt{k^2+m^2-{m^4\over4\k^2}},
\ee
from which it follows that also in this case positive and negative energy states have different masses,  
$m^\pm_0=m\left(-{1\over2\k}\pm\sqrt{1-{m^2\over4\k^2}}\right) $.

In the case of the relativistic hydrogen atom, the KG equation becomes
\be
\left[-\left(p_0-\ar\right)^2+p_i^2+m^2\left(1+\po\right)\right]\f=0,
\ee
In spherical coordinates, using the ansatz \eqref{ansatz}, to first order in $1/\k$ the equation reduces to
\bea
&&\bigg[{\de^2\over\de r^2}+\left(1-\Eo\right){2\over r}{\de\over\de r}-\left(1-2\Eo\right){l(l+1)-\a^2\over r^2}
+\left(1-\Eo\right){2\a E\over r}\cr
&&\quad+~ E^2-m^2\left(1+\Eo\right)\bigg]\q_{lm}=0,
\eea
which has still solutions of the form \eqref{solution}, where $\L$ and $n$ are the same as in \eqref{D-L}-\eqref{eigen} but $\D$ is now
$\D=\sqrt{m^2-E^2+m^2\Eo}$.

One can again expand around the unperturbed value $E_{(0)}$ of the energy \eqref{spec}, obtaining finally
\be
E^{nl}=E_{(0)}^{nl}\left[1\pm{mN\over\k\sqrt{N^2+a^2}}\left(1-{2N^2\over N^2+\a^2}\right)\mp{\a^2\over(N^2+\a^2)}\
{l(l+1)-\a^2\over\sqrt{(l+\ha)^2-\a^2}}\right].
\ee
The characteristics of the spectrum are similar to those found in the previous case.




\section{Outlook and discussion}

In this paper, we have considered $\kappa$-deformed relativistic quantum phase space and the possible implementations of the Lorentz algebra. A first one is a simple extension of the phase space with Lorentz algebra that leaves the Poincar\'e algebra unaltered. Some explicit examples of this implementation have been presented. Another one is a general extension with Lorentz algebra, where the Poincar\'e algebra is deformed. Examples of interpolation between left- and right-covariant realizations corresponding to Poincar\'e-Weyl algebra as well as to the natural realization of the
$\kappa$-Poincar\'e algebra have been  given.

In Sec. 4, we have fixed the Jordanian twist ${\mathcal F}$\eqref{twist-left}, the corresponding realization of noncommutative coordinates $\hat x_\mu$ \eqref{real-left}, the coproduct
of momenta $\Delta p_\mu$ \eqref{deltap-L} and the addition of momenta $(k\oplus q)_\mu=(1+a\cdot q)k_\mu+q_\m$. Then extensions with one-parameter families of realizations of 
Lorentz generators $M^{(u)}_{\mu\nu}$ \eqref{M-u}, dilatations $D^{(u)}$ \eqref{nova} and momentum $p_\mu$ \eqref{p-u} have been considered. They close the Poincar\'e-Weyl
algebra. The commutation relations between $M^{(u)}_{\mu\nu}$ and $D^{(u)}$ and the momentum $p_\mu$ are given in \eqref{poincare}-\eqref{Du-pL}. The corresponding Casimir operator is 
${\mathcal C}=P^2=(p^{(u})^2=p^2/(1+(1-u)a\cdot p)^2$. 

The dispersion relations and their physical consequences have been discussed in the next section, where we have shown how the spectrum of the relativistic hydrogen atom depends on the
representation chosen for the Lorentz algebra.The deviations from special relativity are very small, but are important in principle. An interesting property is the breakdown of the symmetry between positive and negative energy states, due to the peculiar form of the dispesion relations.

\section*{Acknowledgements}
This work was supported in part by the Action MP1405 QSPACE from the European Cooperation in Science and Technology (COST). R.~\v{S}. would like to thank LPT Orsay for the hospitality during her visit.

\end{document}